\documentclass[twocolumn,prb]{revtex4}
\preprint{condmatt}
\usepackage{graphicx}
\usepackage{dcolumn}
\usepackage{bm}
\usepackage{amssymb}

\begin{document}

\title{Failure of Scattering Interference in the Pseudogap State of
  Cuprate Superconductors }

\author{S. Misra}
\author{M. Vershinin}
\author{P. Phillips}
\author{A. Yazdani}
\email{ayazdani@uiuc.edu}
\affiliation{Department of Physics and Fredrick Seitz Materials
  Research Laboratory, University of Illinois at Urbana-Champaign,
  Urbana, Illinois 61801}
\date{\today}

\begin{abstract}
We calculate scattering interference patterns for various electronic 
states proposed for the pseudogap regime of the cuprate
superconductors. The scattering interference models all produce
patterns whose wavelength changes as a function of energy, 
in contradiction to the energy-independent wavelength
seen by scanning tunneling microscopy (STM) experiments in the pseudogap
state. This suggests that the patterns seen in STM local density of
states measurements are not due to scattering interference, but are
rather the result of some form of ordering.

\end{abstract}

\maketitle

The origin of pseudogap phenomena is one of the key questions in the
cuprate superconductors, as it profoundly affects their properties
above the superconducting transition, and potentially influences
the cuprate phase diagram. \cite{Timusk} 
In a recent paper, we have used a scanning tunneling microscope to
show that the electronic states in the pseudogap state of
$Bi_2Sr_2CaCu_2O_{8+\delta}$ are spatially modulated. \cite{Vershinin}
These electronic modulations in the local density of states, 
which are seen only for  
energies less than the pseudogap energy scale, are oriented along the
copper-oxide bond direction and have an energy-independent,
incommensurate wavelength. One interpretation of these experiments is
that these modulations are caused by electronic ordering, 
a variety of which have been proposed for the cuprates in
the pseudogap regime. \cite{SachdevOrder, Eduardo,  Demler,
Zhang, DHLOrder} Another possibility is that such modulations are the
consequence of scattering interference from one of the
electronic states proposed for the pseudogap. \cite{Franz, Norman} 
The interest in scattering interference as a possible explanation is
motivated by the success of this idea in understanding similar
modulations in the superconducting state \cite{Hoffman, DHLee,
McElroy}, although even below $T_c$ 
there are a number of deviations from the
scattering interference picture. \cite{Eduardo,Howald,Vortex,
  DavisOrder} In this paper, we elaborate on calculations
reported in Ref. 2 to demonstrate the failure of the scattering
interference scenario in describing the modulations observed in the
pseudogap state. We show that this failure is a generic feature of
scattering interference itself, regardless of the model chosen for the
pseudogap electronic state. \cite{Franz, ArcFit, Levin, DDW, MFL}
This shortcoming suggests that the modulations observed
in the pseudogap regime are the signature of some form of
ordering.

The scattering interference picture ascribes spatial modulations in
the local density of states to interference between elastically
scattered quasiparticles. \cite {Balatsky, DHLee} A well-studied
example of this phenomenon is
the standing wave patterns seen on certain metal surfaces, in which
electrons scattering from step edges and point
impurities interfere coherently to form modulations in the density of
states. \cite {Eigler} The period of these modulations at a given
energy $\omega$ is determined
by the wavevector $\vec{q}$ that joins the two points on curves of constant
electron energy in $\vec{k}$ space with the greatest weight, as
determined by the electronic structure of the system. 
For a superconductor, the scattering interference contribution to 
the local density of states can be treated within the Born approximation, 
which introduces a correction,
\begin{widetext}
\begin{eqnarray}
\delta n(\vec{r},\omega)=-{1\over\pi} Im\{\int{d^2\vec{r_1}
G_0(\vec{r}-\vec{r_1},\omega) V(\vec{r_1},\omega)
G_0(\vec{r_1}-\vec{r},\omega)-
F_0(\vec{r}-\vec{r_1},\omega) V(\vec{r_1},\omega) 
F_0(\vec{r_1}-\vec{r},\omega)
}\}
\end{eqnarray}
\end{widetext}
 to the density of states, 
where $G_0$ and $F_0$ are the single particle and anamolous Green
functions, respectively, and $V$ is
a weak, finite-range scattering potential. \cite{Franz,DHLee,
Capriotti, Sachdev, Yeh, Hirschfeld} For pure potential scattering,
the Fourier transform of the Born correction 
$\delta n(\vec{q},\omega)=-{1\over\pi}Im\{ 
V(\vec{q},\omega)\Lambda(\vec{q},\omega)\}$
seperates into a part  
\begin{eqnarray}
\Lambda(\vec{q},\omega)=\int\{d^2\vec{k}
G_0(\vec{k},\omega)G_0(\vec{k}+\vec{q},\omega)- \nonumber \\
F_0(\vec{k},\omega)F_0(\vec{k}+\vec{q},\omega)\}%
\label{lambda}
\end{eqnarray}
which contains all the wave interference information, and a part
\begin{equation}
V(\vec{q},\omega) = \int{d^2\vec{x} e^{-i\vec{q}\cdot\vec{x}}
  V(\vec{x},\omega)}
\label{SF}
\end{equation}
which acts like a static structure factor. \cite{Capriotti} 
For now, we will assume that the structure factor does not filter any
wave interference information. 

\begin{figure}[!th]
\begin{center}
\includegraphics[width=8cm]{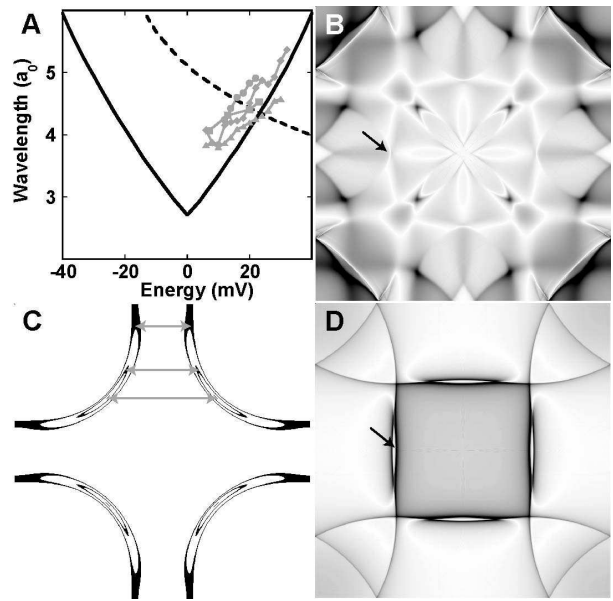}
\caption{\label{SC} (A) STM data in the superconducting state 
  shows modulations in the local density of states which
  disperse. Shown here is data taken along the $(0,\pi)$ direction 
  (the Cu-O bond direction)
  from Vershinin {\it et al.} \cite{Vershinin} at 40K for 
  slightly underdoped $Bi_2Sr_2CaCu_2O_{8+\delta}$ (sqaures), 
  and from Hoffman {\it et al.} \cite{Hoffman}  
  at 4K for underdoped (triangle), as-grown (diamond) and
  overdoped (circle) samples. The wavelength is in units of
  $a_0=3.8\AA$, the Cu-Cu distance. The solid black line shows
  the dispersion for the $(0,\pi)$ mode in the superconducting state
  calculated using Born scattering. The dashed black line shows the
  dispersion of the $(0,\pi)$ mode for the Fermi liquid normal state. 
  (B) The power spectrum of $\delta n(\vec{q},\omega)$ 
  in the superconducting state with a
  broadening of 1mV is shown here in the first Brillouin zone
  at an energy of  $\omega = 20 mV$. Black regions correspond to
  a large contribution from scattering interference. 
  The dispersion of the slowest
  dispersing feature along the
  $(0, \pi)$ direction, highlighted by the arrow, is plotted as the
  solid black line in (A). The $F_0$ term in Eq. \ref{G_sc}
  makes the $(\pi,\pi)$ mode have sharp peaks, but makes the $(0,\pi)$
  mode weaker for non-magnetic impurities. The situation is reversed 
  for the magnetic impurity case (not shown), in which the $F_0$ term has
  the opposite sign. \cite{Franz} (C) The solid black lines show
  contours of constant electron energy in the superconducting state 
  at $\omega = 0mV$ (point along
  $(\pi,\pi)$ direction), $\omega = \Delta/2$ (small banana) 
  and $\omega = \Delta$ (large banana). The most
  prominent peaks in scattering wavevector space (q-space) along the
  $(0,\pi)$ direction are indicated by the grey arrows. 
  (D) The power spectrum of
  $\delta n(\vec{q},\omega)$ in the Fermi liquid normal state with a
  broadening of 1mV is shown here in the first Brillouin zone
  at an energy of  $\omega = 20 mV$. 
  The dispersion of the slowest dispersing feature along the 
  $(0, \pi)$ direction, highlighted by the arrow, is plotted as the
  dashed black line in (A).}
\end{center}
\end{figure}

In order to develop a context in which to understand quantum
interference in the pseudogap state, we will first review the quantum
interference picture for the two cases already discussed in the
literature- the superconducting state and the Fermi liquid normal state. 
\cite{DHLee, Capriotti, Sachdev, Hirschfeld} Following these previous
works, scattering interference in the superconducting state can be
modeled using the Green functions, 
\begin{eqnarray}
G_0(\vec{k},\omega)={\omega+i\delta+\epsilon_{\vec{k}}\over
{(\omega+i\delta-\epsilon_{\vec{k}})
(\omega+i\delta+\epsilon_{\vec{k}})-{\Delta_{\vec{k}}^2}}} \nonumber\\
F_0(\vec{k},\omega)= {\Delta_{\vec{k}}\over
{(\omega+i\delta-\epsilon_{\vec{k}})
(\omega+i\delta+\epsilon_{\vec{k}})-{\Delta_{\vec{k}}^2}}}
\label{G_sc}
\end{eqnarray}
where 
$\epsilon_{\vec{k}}=120.5 - 595.1\times(\cos{k_x}+\cos{k_y})/2 + 
163.6\times\cos{k_x}\cos{k_y} - 51.9\times(\cos{2k_x}+\cos{2k_y})/2 -
111.7\times(\cos{2k_x}\cos{ky}+\cos{k_x}\cos{2k_y})/2 +
51.0\times(\cos{2k_x}\cos{2k_y})$ 
is the $Bi_2Sr_2CaCu_2O_{8+\delta}$ band structure from ARPES
\cite{BS} for slightly underdoped ($\delta=.12$) samples,
$\Delta_{\vec{k}}=45.0\times(\cos{k_x}+\cos{k_y})/2$ is the
superconducting gap function, and $\delta$ is a broadening term.
As shown in Figures \ref{SC}a and b, this yields pictures with sharp
peaks in q-space which disperse. To understand the origin of the
dispersion qualitatively, it helps to consider a simplified version of 
scattering interference, the so-called octet model,
introduced by Hoffman {\it et al.} and McElroy {\it et al.}. 
\cite{Hoffman, McElroy} Quasiparticles can
elastically scatter between points on contours of constant electron
energy in k-space (Fig. \ref{SC}c). The shape of these
contours changes continuously as the energy increases.
Consequently, the length of the characteristic
scattering interference wavevector (q-space) changes, leading to
dispersion. This picture changes dramatically if, instead of a superconducting
Green function, we use a Fermi liquid normal state Green function by 
setting $\Delta_{\vec{k}}=0$ in Equation 4. \cite{Capriotti}
The sharp peaks in q-space for the
superconducting state have been replaced with dispersing caustics
(Figs. \ref{SC}a and d). The equivalence of all points in
 k-space leads directly to an absence of sharp peaks in q-space. 
The dispersion is now a direct consequence 
of the band structure itself. Although the
Fermi liquid scattering interference picture is useful in determining the
effect of the band structure on scattering interference, it is not
applicable to the pseudogap, as it omits two key features of
this state: the pseudogap in the density of states, 
and the ill-defined nature of quasiparticles.

\begin{figure}[!bh]
\begin{center}
\includegraphics[width=8cm]{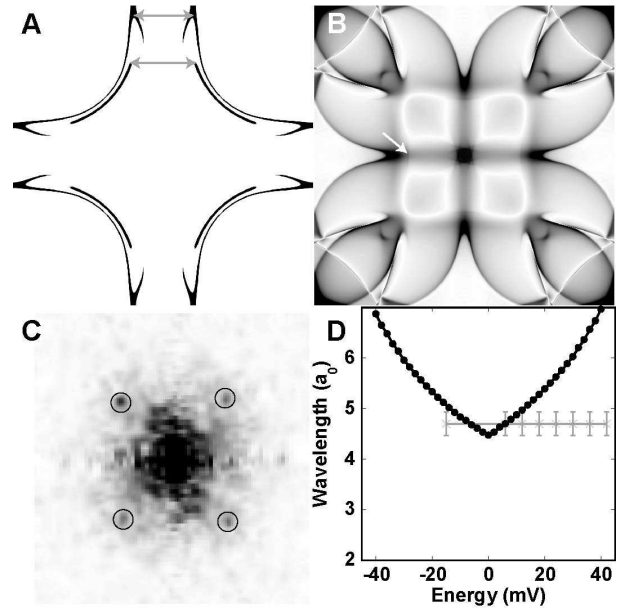}
\caption{\label{Arcs} (A) Shown in black are model representations of
  the electronic density of states in k-space measured by ARPES 
  at the Fermi energy (arcs centered at the nodal
  points) and at the pseudogap energy (banana). 
  Shown in gray are the shortest wavevectors along the
  $(0,\pi)$ direction for scattering interference at the two respective
  energies. (B) We calculated the power spectrum of $\delta 
  n(\vec{q},\omega)$ for a  phenomenological 
  Fermi arc Green function where the 
  extent of the Fermi
  arcs were chosen to give a scattering wavelength
  of $4.6 a_0$ at the Fermi energy, and unphysically small values
  of $\Gamma_0=1mV$ and $\Gamma_1=1mV$ were chosen to produce sharper
  features. The  power spectrum of $\delta  n(\vec{q},\omega)$ 
  for this Fermi arc picture is shown here in the
  first Brillouin zone at $\omega = 20mV$. The dispersion of the
  most slowly dispersing 
  feature along the $(0,\pi)$ direction, indicated by the arrow, 
  is shown in (D) as a
  solid line. (C) The power spectrum of
  the STM data at 15mV in the pseudogap state from Vershinin {\it et al.}
  \cite{Vershinin} shows four peaks along the
  $(0,\pi)$ direction (circles). We have magnified
  the data to show only the relevant region in
  q-space. (D) These peaks
  in q-space correspond to a fixed wavelength of $4.7 \pm 0.2 a_0$,
  shown here as the grey line. Also shown is the dispersion of
  the most slowly dispersing feature along the $(0,\pi)$ direction in 
  scattering interference calculations for the Fermi arc Green function. 
  This feature disperses 
  similar to the $(0,\pi)$ mode in the superconducting state for small
  ($\approx 1mV$)
  values of $\Gamma_0$ and $\Gamma_1$. Larger ($\approx 10mV$)
  values of $\Gamma_0$
  lead to the gradual suppression of this mode in favor of 
  a mode which disperses similar to the
  band structure (Fig. \ref{SC}a). Larger values of
  $\Gamma_1$ yield pictures which average over some part of the
  dispersion for a particular mode, leading to unphysically broad features.
}
\end{center}
\end{figure}

To extend the scattering interference model into the pseudogap regime,
we  must first  choose  an  appropriate Green  function with which
to model  the
pseudogap state. Although its origin is still not understood, the pseudogap
has been thoroughly characterized by ARPES in the underdoped cuprate
superconductors. \cite{Arcs} At all energies, peaks in the ARPES
spectral function are so broad that it 
is unclear whether quasiparticles are still well-defined. \cite{SAG} 
Meanwhile, at the Fermi energy, the spectral function shows
extended arcs centered at the nodal points (Fig. \ref{Arcs}a) in k-space. 
As energy increases towards the pseudogap energy scale, these arcs
extend towards the Brilouin zone boundary. 
Theoretical attempts at understanding the measured 
behaviour have either followed a phenomenological approach, or have
proposed exotic electronic states for the pseudogap. We will first focus on 
the phenomenological attempts. Norman {\it et al.} have modeled the 
ARPES data in the pseudogap regime using the Green function
\begin{equation}
G_0(\vec{k},\omega)=(\omega-\epsilon_{\vec{k}}-\Sigma_{\vec{k}})^{-1}
\label{SE}
\end{equation}
and $F_0=0$,
where the self energy,
\begin{equation}
\Sigma(\vec{k},\omega)=-i\Gamma_1+\Delta_{\vec{k}}^2/
(\omega+\epsilon_{\vec{k}}+i\Gamma_0),
\label{Norm}
\end{equation}
$\Gamma_1$ is a single
particle scattering rate, $\Gamma_0$ is a measure of decoherence, and
$\Delta_{\vec{k}}$ is a gap function with a $\vec{k}$ dependence
which matches the Fermi arcs. \cite{ArcFit} 
The calculated scattering interference patterns (Fig. \ref{Arcs}b)
have two notable shortcomings in comparison with the measured data shown
in Figure \ref{Arcs}c. First,  the calculated patterns contain caustics,
while  the  measured  patterns   contain  discrete  points  along  the
$(0,\pi)$ directions.  This shortcoming can be overcome by
assuming, for example, that the STM tunneling matrix element has a
$d$-wave symmetry and selectively filters the density of states 
along the $(0, \pi)$ direction. 
Second, the scattering interference patterns 
calculated from this phenomenological description
still show features along the $(0,\pi)$ directions that 
disperse (Fig. \ref{Arcs}d). This dispersion curve resembles that in the
superconducting state, as the characteristic wavevector for
scattering interference is of a similar length at the gap energy, but is
shallower because this wavevector is much shorter at the Fermi energy
in the pseudogap regime (Figs. \ref{SC}c and  \ref{Arcs}a). Even
this shallower dispersion results in a change of wavelength between $7
a_0$ and $4.6 a_0$ between the Fermi energy and the pseudogap energy,
whereas STM experiments show a fixed wavelength of $4.7 \pm 0.2 a_0$
over this energy range. \cite{Vershinin} 

\begin{figure}[!th]
\begin{center}
\includegraphics[width=8cm]{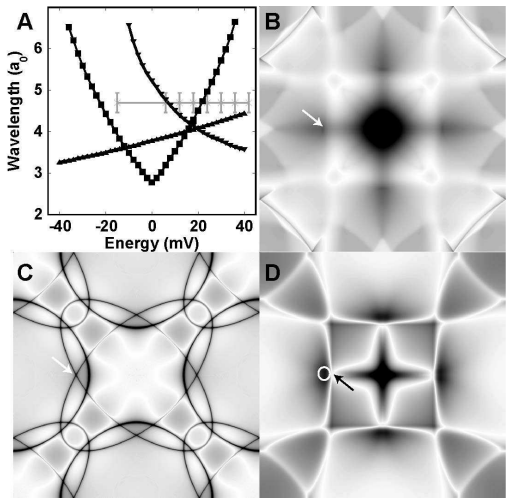}
\caption{\label{Other} (A) The dispersions of all the exotic
  electronic state Green functions discussed in the text should be
  resolved by the STM experiment. Shown are the dispersions of the
  most slowly dispersing
  $(0,\pi)$ modes for the $QED_3$ (squares), DDW (up triangles) and
  circulating current (down triangles) Green functions. These modes
  are labeled by the arrows in parts (B), (C), and (D). STM data from the
  pseudogap state \cite{Vershinin} is shown in gray. (B) The
  power spectrum of $\delta 
  n(\vec{q},\omega)$ for a $QED_3$ Green function 
   is shown here in the
  first Brillouin zone at $\omega = 20mV$. The
  patterns were calculated using a value of $\eta =0.4$ and a
  broadening of 1mV. For
  increased $\eta$, the patterns in q-space become more diffuse until,
  by $\eta\approx1$, there are no discernable features in q-space. (C) The
  power spectrum of $\delta 
  n(\vec{q},\omega)$ for a DDW Green function using a broadening of 1mV
  and a chemical potential
  shift of $+40 meV$ is shown here in the
  first Brillouin zone at $\omega = 20mV$. (D) The
  power spectrum of $\delta 
  n(\vec{q},\omega)$ for a circulating current 
    Green function using $\lambda=.27$, $T=10mV$, $D_0=10mV$, and 
  $\omega_c=500mV$ 
   is shown here in the
  first Brillouin zone at $\omega = 20mV$. Non-zero values of $D_0$
  result in the slowest dispersing mode along the $(0,\pi)$ direction
  (arrow) being replaced by a faster dispersing mode (circle)
  at energies $\omega<|2\times D_0|$. The dispersion shown in part (A)
  is for $D_0=0$, corresponding to the slowest dispersing mode.}
\end{center}
\end{figure}

We next consider the scattering interference scenario for various exotic
electronic states proposed for the pseudogap regime. 
We will discuss how four of these exotic Green
functions (two based on preformed pairs, and two  based on
ordering) produce dispersions which should be resolved in the data.
Chen {\it et al.} \cite{Levin} have proposed the Green function in
Equation \ref{SE} with the self energy in Equation \ref{Norm} 
to describe preformed pairs. However, unlike the Fermi arc model, they use a
$d$-wave gap function. For small values of $\Gamma_0$,
the scattering pictures resemble those calculated for the
superconducting state, and for larger values, they resemble broader
versions of those calculated for the Fermi liquid (not shown). Both
these dispersions, similar to the ones shown in Figure 1a, should be
resolved by the experiment.
Another example of a preformed pairs
calculation appears in Pereg-Barnea and Franz \cite{Franz}, 
where the authors take a
$QED_3$ Green function $G_0(\vec{k},\omega)=
(\omega+\epsilon_{\vec{k}})/
(\omega^2+\epsilon_{\vec{k}}^2+\Delta_{\vec{k}}^2)^{1-\eta/2}$, with
$\eta$ being an anomolous dimension exponent which
increases the amount of decoherence of the patterns for larger
values. As noted by the authors, the scattering
interference patterns show a dispersion identical to the superconducting
state for $\eta<0.5$, which is shown here explicitly in Figures
\ref{Other}a and b. 
We have also calculated the dispersion
for a simplified d-density wave (DDW) ordering Green function $G_0(\vec{k},
\omega)= (\omega+\Delta\mu-
\epsilon_{\vec{k}+\vec{Q}})/((\omega-\Delta\mu-\epsilon_{\vec{k}}) 
(\omega-\Delta\mu-\epsilon_{\vec{k}+\vec{Q}})-D_{\vec{k}}^2)$, where
$D_{\vec{k}}=D_0 
(\cos{k_x}-\cos{k_y})/2$ is the DDW gap, $\Delta\mu$ is a chemical
potential shift,  and $\vec{Q}=(\pi,\pi)$. 
\cite{Franz,DDW}
As can be seen in Figures \ref{Other}a and c, the simplified 
DDW Green function
produces scattering interference patterns which disperse through a
range of wavelengths ($\Delta \lambda=0.8a_0$ between -15mV and 35mV)
which should be easily
resolved by the STM experiment (maximum $\Delta \lambda=0.4a_0$ over
the same range of energies). Using a more complete DDW
  treatment \cite{Franz, DDW} results in an identical dispersion, and
  very similar scattering pictures (not shown).  Finally, in
Figures \ref{Other}a and d, we show the quantum
intereference patterns for the marginal Fermi liquid scenario with
circulating current order.
\cite{MFL} The Green Function for this phase, given by
$G_0(\vec{k},\omega)=(\omega-\epsilon_{\vec{k}} \pm D(\vec{k})
-\Sigma_{\vec{k}})^{-1}$ for $k \gtrless k_f$
with $\Sigma(\vec{k},\omega)=
\lambda\{\omega \log{x/\omega_c}+i\pi x/\cosh{D(\vec{k})/x}\}$, 
$x=(\omega^2+\pi^2T^2)^(1/2)$ and
$D(\vec{k})=D_0(\cos(k_xa)-\cos(k_ya))^2$, produces
patterns which disperse in a fashion similar to the band
structure. Thus, it also fails
to match the dispersionless feature seen in STM in the pseudogap regime
(Fig. \ref{Other}a). 

The failure of these approaches points to a fundamental problem:
any Green function 
will result in a wave contribution to Born scattering which
disperses. The Green function is being asked to
accomplish two contradictory tasks: on one hand, the imaginary part of
the Green function (the density of states) must disperse in order to
match 
the ARPES band structure, and on the other hand, the imaginary part
of the Green function convolved with itself cannot disperse if it is
to match the STM data in the pseudogap state. 
In order for Born scattering to simultaneously match the ARPES and STM
data, the structure factor, assumed to be an all-pass filter until
now, could be chosen to pass only contributions
near $\vec{q}=2\pi/a_0(0,1/4.7)$ inside the pseudogap. \cite{Capriotti}
The real space scattering potential 
corresponding to such a structure factor is
an incommensurate,  bond-oriented square lattice of scattering
centers.  While this 
can be justified on various physical grounds \cite{Sachdev, Yeh}, 
it amounts to an {\it ad hoc} assumption that ordering
exists. Moreover, any choice of Green function would then adequately
describe the data, thus demonstrating that scattering interference fails
to describe the interesting physics revealed in the experiment. 

\begin{acknowledgments}
We wish to acknowledge helpful conversations with Mohit Rendaria and
Mike Norman. Work supported
under NSF (DMR-98-75565, DMR-0305864, \& DMR-0301529632), 
DOE through Fredrick Seitz
Materials Research Laboratory (DEFG-02-91ER4539), ONR (N000140110071),
Willet Faculty Scholar Fund, and Sloan Research Fellowship. AY
acknowledges support and hospitality of D. Goldhaber-Gordon and
K.A. Moler at Stanford. 
\end{acknowledgments}

\bibliography{apssamp}

\begin{references}

\bibitem{Timusk} T. Timusk and B.W. Statt, Rep. Prog. Phys. {\bf 62},
  61 (1999). 
\bibitem{Vershinin} M. Vershinin {\it et al.}, Science {\bf 303}, 1996
  (2004). 
\bibitem{SachdevOrder} S. Sachdev, Rev. Mod. Phys. {\bf 75}, 913 (2003).
\bibitem{Eduardo} S.A. Kivelson {\it et al.}, Rev. Mod. Phys. {\bf
    75}, 1201 (2003).
\bibitem{Demler} D. Podolsky, E. Demler, K. Damle, and B.I. Halperin,
  Phys. Rev. B {\bf 67}, 94514 (2003).
\bibitem{Zhang} H.-D. Chen, O. Vafek, A. Yazdani, and S.-C. Zhang,
  cond-matt/0402323 (2004).
\bibitem{DHLOrder} H.C. Fu, J.C. Davis, and D.-H. Lee, cond-mat/0403001
  (2004). 

\bibitem{Norman} M. Norman, Science {\bf 303}, 1985 (2004).
\bibitem{Franz} T. Pereg-Barnea and M. Franz, Phys. Rev. B {\bf 68}
  180506 (2003). 

\bibitem{Hoffman} J.E. Hoffman {\it et al.}, Science {\bf 297}, 1148
  (2002).
\bibitem{McElroy} K. McElroy {\it et al.}, Nature {\bf 422}, 592
  (2003). 
\bibitem{DHLee} Q.-H. Wang and D.-H. Lee, Phys. Rev. B {\bf 67} 020511
  (2003). 

\bibitem{Howald} C. Howald, H. Eisaki, N. Kaneko, M. Greven, and
  A. Kapitulnik, Phys. Rev. B {\bf 67}, 014533 (2003).
\bibitem{Vortex} J.E. Hoffman {\it et al.}, Science {\bf 295}, 466
  (2002). 
\bibitem{DavisOrder} K. McElroy, {\it et al.}, cond-mat/0404005
  (2004).

\bibitem{ArcFit} M.R. Norman, M. Randeria, H. Ding, and J.C. Campuzano,
  Phys. Rev. B {\bf 57}, 11093 (1998). 
\bibitem{Levin} Q. Chen, K. Levin, and I. Kosztin, Phys. Rev. B {\bf
    63}, 184519 (2001).
\bibitem{DDW} C. Bena, S. Chakravarty, J. Hu, and C. Nayak, cond-mat/
  0311299 (2003). 
\bibitem{MFL} C. M. Varma, Phys. Rev. Lett. {\bf 83}, 3538 (1999).

\bibitem{Balatsky} M.I. Salkola, A.V. Balatsky, and D.J. Scalapino,
  Phys. Rev. B {\bf 77}, 1841 (1996).
\bibitem{Eigler} M.F. Crommie, C.P. Lutz, and D.M. Eigler, Nature {\bf
    363} 524 (1993).

\bibitem{Capriotti} L. Capriotti, D.J. Scalapino, and R.D. Sedgewick,
  Phys. Rev. B {\bf 68} 014508 (2003).
\bibitem{Sachdev} A. Polkovnikov, M. Vojta, and S. Sachdev,
  Physica C {\bf 388-389}, 19 (2003).
\bibitem{Yeh} C.-T. Chen and N.-C. Yeh, Phys. Rev. B {\bf 68} 220505
  (2003). 
\bibitem{Hirschfeld} L. Zhu, W.A. Atkinson, and P.J. Hirschfeld,
  Phys. Rev. B {\bf 69}, 060503 (2004).
\bibitem{BS} M.R. Norman, M. Randeria, H. Ding, and J.C. Campuzano,
  Phys. Rev. B {\bf 52}, 615 (1995).
\bibitem{Arcs} M.R. Norman, {\it et al.}, Nature {\bf 392}, 157 (1998).
\bibitem{SAG} T. Valla {\it et al.}, Science {\bf 285}, 2110 (1999).
\end{references}
\end{document}